\documentclass[a4paper,10pt]{article}
\usepackage{
times,xcolor,
amsmath,amssymb,epsfig,graphics,graphicx}

\newcommand{\be}{\begin{eqnarray}}
\newcommand{\ee}{\end{eqnarray}}

\begin{document}

\begin{center}
{\Large\bf {Second order differential equations for bosons with spin $j\geq 1$ and in the bases of  general tensor-spinors of rank-$2j$
}}
\end{center}
\vspace{0.02cm}

\begin{center}
 V.\ M.\ Banda Guzm\'an and M.\ Kirchbach
\end{center}

\vspace{0.01cm}
\begin{center}
{\it Instituto de F\'{\i}sica},
         {\it Universidad Aut\'onoma de San Luis Potos\'{\i}},\\
         {\it Av. Manuel Nava 6, San Luis Potos\'{\i}, S.L.P. 78290, M\'exico}
\end{center}

\vspace{0.01cm}

\begin{flushleft}{{\bf Abstract}:
A boson of spin-$j\geq 1$ can be  described in one of the possibilities within the Bargmann-Wigner framework by means of one sole differential equation of  order twice the
spin, which however  is known to be inconsistent as it allows for non-local, ghost and acausally propagating solutions, all problems which are difficult to tackle. The other possibility is provided by the Fierz-Pauli framework which  is based on the  more comfortable to deal with  second order Klein-Gordon equation, but it needs to be supplemented by an auxiliary condition. Although the latter formalism avoids  some of the pathologies  of the high-order equations, it still remains plagued by some inconsistencies such as the  acausal propagation of the wave fronts of the (classical) solutions within an electromagnetic environment. We here suggest a method  alternative to the above two  that  combines their advantages while avoiding the related difficulties. Namely, we suggest one sole strictly  $D^{(j,0)\oplus (0,j)}$ representation specific  second order differential equation, which is derivable from a Lagrangian and whose solutions do not violate causality. The equation under discussion presents itself as  the product of the Klein-Gordon operator with a momentum independent projector on Lorentz irreducible representation spaces constructed from one of the Casimir invariants of the spin-Lorentz group. The basis used is that of  general tensor-spinors of rank-$2j$.

}
\end{flushleft}

\begin{flushleft}
\end{flushleft}
\tableofcontents

\section{Introduction.} High spin $j\geq 1$ fields, both massive and massless, have always been among the principal topics in field theories. In particle physics they  are needed in the description of the reported hadron resonances whose spins vary from $1/2$ to $17/2$ for baryons, and from zero to six for mesons. In gravity, bosons of higher spins can couple to the metric tensor and cause its deformation \cite{Gravity}, and are besides this in demand in the physics of rotating black holes \cite{rot_BH}. Gravitational interactions among high-spin fermions are also under discussion \cite{A}. Various approaches to  high-spin fields in general and bosonic in particular have been developed over the years (see \cite{HSReview} for a recent review, and
\cite{BR} for a standard textbook),  the Fierz-Pauli (FP) \cite{PF} and the Bargmann-Wigner (BW) frameworks \cite{BW1948} counting  among them. 
In the following we briefly highlight these two methods for the sake of self consistency of the presentation, comment on their problems, and suggest an alternative approach to high-spin description  based on one sole  strictly  representation specific second order differential equation which is free from inconsistencies.

\subsection{General tensor-spinors and their restrictions within the  Bargmann-Wigner and Fierz-Pauli frameworks for 
spin-$j\geq 1$. }
The concepts underlying the two methods under discussion are most transparently  presented within the framework of spinor calculus \cite{Spinorbook}, relevant for gravity in the spinor form.
Specifically, both methods depart from general tensor-spinors, here denoted by $\left(S^r_s\right){}^{\alpha_1...\alpha_r}{}{}{}_{\stackrel{\bullet}{\beta}_1...\stackrel{\bullet}{\beta}_s}$, and defined as products of $r$   spinor components, ( $\xi^{\alpha_i}$),   with   $s$ co-spinors components, ($\eta_{\stackrel{\bullet}{\beta}_k}$) according to,
\begin{equation}
\left( S^{r}_{s}\right) {}^{\alpha_1...\alpha_r}{}{}{}_{\stackrel{\bullet}{\beta}_1...\stackrel{\bullet}{\beta}_s}=
\stackrel{(1)}{\xi^{\alpha_1}}...\stackrel{(r)}{\xi^{\alpha_r}}\stackrel{(1)}{\eta_{\stackrel{\bullet}{\beta}_1}}...\stackrel{(s)}{\eta_{\stackrel{\bullet}{\beta}_s}}.
\label{spin-tensor}
\end{equation}
Here the marks in the parentheses on top of the spinor components enumerate the  different $sl(2,C)$ spinors to which they belong.
{}For the sake of not overloading the notations, these marks will be suppressed in the following. 
The Bargmann-Wigner (BW) method specifically singles out  the  tensor-spinors, $S^n_0$ and $S^0_n$,  
\begin{eqnarray}
\mbox{BW}:\quad \left(S^n_0\right)^{\alpha_1...\alpha_n}=\xi^{\alpha_1}...\xi^{\alpha_n}, &\quad& 
\left( S^0_n\right)_{\stackrel{\bullet}{\beta_1}...\stackrel{\bullet}{\beta_n}}=\eta_{\stackrel{\bullet}{\beta_1}}...\eta_{\stackrel{\bullet}{\beta_n}},\nonumber\\
\end{eqnarray}
embedded within the general direct product, 
\begin{eqnarray}
{ S}^{r}_s \simeq \otimes _{i=1}^{i=n}\, \left[D^{(1/2,0)\oplus (0,1/2)}\right]_i,\quad D^{(1/2,0)\oplus (0,1/2)}=
\left(
\begin{array}{c}
\xi^1\\
\xi^2\\
\eta_{\stackrel{\bullet}{1}}\\
\eta_{\stackrel{\bullet}{2}}
\end{array}
\right),\quad r+s=n,
\label{Gl_prm}
\end{eqnarray}
of $n$ four-component  spinors, $ D^{(1/2,0)\oplus (0,1/2)}$.
Instead, the Fierz-Pauli framework is based on rank-$2n$ tensor-spinors, $S^n_n$, (subsequently denoted by $\Phi$) which have an equal number of undotted (spinor) and dotted(co-spinor) indexes, namely,
\begin{eqnarray}
\mbox{FP}: \hspace{2.1cm}{\Phi}^{\alpha_1}_{\stackrel{\bullet}{\beta_1}}{}^{...}_{...}{}{}^{\alpha_n}_{\stackrel{\bullet}{\beta_n}}&=&
\xi^{\alpha_1}\eta_{\stackrel{\bullet}{\beta_1}}...\xi^{\alpha_n}\eta_{\stackrel{\bullet}{\beta_n}}.
\end{eqnarray} 
They emerge  as the direct product of $n$ four-vectors, $D^{(1/2,1/2)}$, as
\begin{eqnarray}
\mbox{BW}:\qquad \Phi&\simeq& \otimes _{i=1}^{i=n}\,\left[ D^{(1/2,1/2)}\right]_i, \quad D^{(1/2,1/2)}=\left(
\begin{array}{cc}
\xi^{1}\eta_{\stackrel{\bullet}{1}}&\xi^{1}\eta_{\stackrel{\bullet}{2}}\\
\xi^{2} \eta_{\stackrel{\bullet}{1}}&\xi^{2}\eta_{\stackrel{\bullet}{2}}
\end{array}
\right).
\label{Phispace}
\end{eqnarray}
The  ${\mathcal S}^{r}_{s}$ and $\Phi$ spaces in the respective equations (\ref{Gl_prm}) and (\ref{Phispace}) decompose into irreducible Lorentz group representation spaces as,
\begin{eqnarray}
\mbox{BW}:\quad \otimes _{i=1}^{i=n}\, \left[D^{(1/2,0)\oplus (0,1/2)}\right]_i&=&D^{\left(0,0\right)}\oplus ... \oplus
D^{\left(\frac{n}{2},0\right)}\oplus D^{\left(0,\frac{n}{2}\right)}\oplus ...\oplus D^{ \left(\frac{1}{2},\frac{n-1}{2} \right)} \oplus
D^{\left(\frac{n-1}{2}, \frac{1}{2} \right)}\oplus...\nonumber\\
&\oplus& D^{\left(m\frac{1}{2},m\frac{1}{2}\right)}\oplus..., \qquad m \leq \frac{n}{4},
\label{Gl1}
\end{eqnarray}
for the BW method, and as
\begin{eqnarray}
\mbox{FP:}\qquad \quad \otimes_{i=1}^{i=n}\left[ D^{(1/2,1/2)}\right]_i&=&
D^{\left(0,0\right)}\oplus ... \oplus  D^{\left(0,\frac{n}{2}\right)}\oplus D^{\left(\frac{n}{2}, 0\right)}\oplus ...
\oplus D^{\left(\frac{n}{2},\frac{n}{2}\right)}, \,\,
\label{FP-tsr-rdct}
\end{eqnarray}
for the FP method.
The equation (\ref{Gl1})  shows that there are several irreducible  Lorentz group representation spaces of different dimensionality  
which contain the highest  possible spin-$j=n/2$, such as the multiple-spin $D^{(n/4,n/4)}$--, the two-spin 
$D^{(1/2,(n-1)/2 )\oplus ((n-1)/2,1/2)}$--, and the pure spin $D^{(n/2,0)\oplus (0,n/2)}$ spaces,  the latter being of main interest to the BW method. On the other side, the highest possible spin in (\ref{FP-tsr-rdct}) is twice as big, given the fact that the spinor-tensor is of rank-$2n$. It is considered as the highest spin $j=n$ in the $D^{\left(\frac{n}{2},\frac{n}{2}\right)}$ irreducible sector of $\Phi$, the former containing  multiple spins $j$ which fall within the range of $j\in\left[0,n \right]$.

In order to identify $ D^{\left(\frac{n}{2}, 0\right)\oplus \left(0,\frac{n}{2}\right)}$ in (\ref{Gl1}), the Bargmann-Wigner approach executes the following strategy:
\begin{itemize}
\item It constructs the  four-spinor of rank-$n$, here denoted by $\psi^{(n)}$ as
\begin{eqnarray}
\psi^{(n)}=\left( 
\begin{array}{c}
S^{n}_{0}\\
S^{0}_{n}
\end{array}
\right) &=&
\left(
\begin{array}{c}
\psi^{1, \alpha_2,...\alpha_{n}}\\
\psi^{2,\alpha_2...\alpha_n}\\
\psi_{\stackrel{\bullet}{1}\stackrel{\bullet}{\beta_2}...\stackrel{\bullet}{\beta_{n}}}\\
\psi_{\stackrel{\bullet}{2}\stackrel{\bullet}{\beta_2}...\stackrel{\bullet}{\beta_{n}}}
\end{array}
\right),\label{Gl2}
\end{eqnarray}

\item It confines to symmetric spinor indexes alone according to,
\begin{eqnarray}
\mbox{Sym}{}\, \psi^{\alpha_1... \lbrace \alpha_i...\alpha_r\rbrace  \dots\alpha_n}&=&\mbox{Sym}{}\, \psi^{\alpha_1... \lbrace \alpha_r...\alpha_i\rbrace  \dots\alpha_n}, \nonumber\\
\mbox{Sym}{}\, \psi_{\stackrel{\bullet}{\beta_1}... \lbrace \stackrel{\bullet}{\beta_i}...\stackrel{\bullet}{\beta_r}\rbrace  \dots\stackrel{\bullet}{\beta _n}}&=&
\mbox{Sym}{}\, \psi_{\stackrel{\bullet}{\beta_1}... \lbrace \stackrel{\bullet}{\beta_r}...\stackrel{\bullet}{\beta_i}\rbrace  \dots\stackrel{\bullet}{\beta_n}}.
\label{SYM}
\end{eqnarray}

\item It constructs the dynamics by requiring the symmetrized rank-$n$  spinor to be eigenstate $\forall i$  to  the covariant projector,
$\Pi^{BW}\left(\partial\right)$, defined as,
\begin{eqnarray}
\Pi ^{BW}\left(\partial^{2j}\right)&=&\otimes _{r=1}^{r=n}\left[ \pi(\partial)\right]_r, \nonumber\\
 \pi(\partial) &=&\frac{{i\mathcal D} +m}{2m},
\label{Gl3}
\end{eqnarray}
with ${\mathcal D}$ being  defined as \begin{eqnarray}
{\mathcal D}&=&
\left(\begin{array}{cccc}
0&0&\partial^{1\stackrel{\bullet}{1}}&\partial^{1\stackrel{\bullet}{2}}\\
0&0& \partial^{2\stackrel{\bullet}{1}}&\partial^{2\stackrel{\bullet}{2}}\\
\partial_{1\stackrel{\bullet}{1}}&\partial_{1\stackrel{\bullet}{2}}&0&0\\
\partial_{2\stackrel{\bullet}{1}}&\partial_{2\stackrel{\bullet}{2}}&0&0                             
\end{array}
\right), 
\end{eqnarray}
where
\begin{eqnarray}
\left(
\begin{array}{cc}
\partial^{1\stackrel{\bullet}{1}}&\partial^{1\stackrel{\bullet}{2}}\\
\partial^{2\stackrel{\bullet}{1}}&\partial^{2\stackrel{\bullet}{2}}
\end{array}
\right)
&=&\partial^0 \sigma_0 +\nabla \cdot {\vec \sigma}, \quad
\left( \begin{array}{cc} 
\partial_{1\stackrel{\bullet}{1}}&\partial_{1\stackrel{\bullet}{2}}\\
\partial_{2\stackrel{\bullet}{1}}&\partial_{2\stackrel{\bullet}{2}}
\end{array}\right)=
\partial^0\sigma_0 -\nabla\cdot \vec{\sigma}^T,\quad
  \sigma_0 =
\left(\begin{array}{cc}
1&0\\
0&1 
\end{array}\right),
\label{Dt_Prc}
\end{eqnarray}
which ensures that the spinor under discussion satisfies the Dirac equation $\forall i$ as:
\begin{eqnarray}
{\mathcal D}\left(
\begin{array}{c} 
\xi^1\\
\xi^2\\
\eta_{\stackrel{\bullet}{1}}\\
\eta_{\stackrel{\bullet}{2}}
\end{array}
\right)= -im\left(
\begin{array}{c} 
\xi^1\\
\xi^2\\
\eta_{\stackrel{\bullet}{1}}\\
\eta_{\stackrel{\bullet}{2}}
\end{array}
\right).
 \end{eqnarray}
\end{itemize}
Stated differently, the Bargmann-Wigner framework \cite{BW1948} for the description of particles with any spin $j$ (preceded by a work of de Broglie \cite{deBroigle} for spin-$1$) is based on the general principle for constructing irreducible representation spaces of the spin-Lorentz group (acting
on the internal spin degrees of freedom alone) as parts of  multiple direct products of  its fundamental four-component spinor, 
$\xi\oplus \stackrel{\bullet}{\eta}$, the direct sum of a spinor, $\xi$,  and co-spinor,$\stackrel{\bullet}{\eta}$, \cite{Spinorbook} 
 as a principal building block.
The BW projector imposes the Dirac equation (and therefore the on-mass shell condition) on each one of the spinor components in the product, a reason for which
it becomes of  the order $\partial^{n}\equiv \partial^{2j}$  in the derivatives, as indicated in the parentheses of its notation. 
With the aid of the explicit expressions of $\partial^{\alpha\stackrel{\bullet}{\beta}}$ , and $\partial_{\alpha\stackrel{\bullet}{\beta}}$, in (\ref{Dt_Prc}) one finds  the Bargmann-Wigner framework expressed in terms of the  following two coupled high-order differential equations,
\begin{eqnarray}
j=n/2\in D^{\left(\frac{n}{2},0\right)\oplus \left(0,\frac{n}{2}\right)}:\quad  \left( \partial^{\alpha_1\stackrel{\bullet}{\beta_1}}\otimes \dots \otimes  \partial^{\alpha_n\stackrel{\bullet}{\beta_n}}\right) 
\mbox{Sym}{}\, \psi_{\stackrel{\bullet}{\beta_1}\dots\stackrel{\bullet}{\beta_n}} &=& (-im)^{n}\mbox{Sym}{}\, \psi{}^{\alpha_1\dots\alpha_n},
\label{BW-1} \\
\left(\partial_{\alpha_1\stackrel{\bullet}{\beta_1}}\otimes \dots \otimes \partial_{\alpha_n\stackrel{\bullet}{\beta_n}}\right)
\mbox{Sym}{}\, \psi^{\alpha_1 \dots\alpha_n}&=&(-im)^n \mbox{Sym}{}\, \psi_{\stackrel{\bullet}{\beta_1}\dots\stackrel{\bullet}{\beta_n}} .
\label{BW-2}
\end{eqnarray}
It can be shown that the BW scheme selects precisely  $2(2j+1)$ degrees of freedom, as required for the description of the
two representation spaces, $D^{(j,0)}$ and $D^{ (0,j)}$ .
In this fashion, spin-$j=n/2$ is described within the BW framework by means of a totally symmetric rank-$2j$ four-component spinor and in terms of the
higher-order differential equations in (\ref{BW-1})-(\ref{BW-2}).
However, as remarked above, such equations present serious difficulties in so far as they allow for unphysical non-local and ghost solutions
\cite{WenliangLi},\cite{TaiJunChen} which need a special effort to be excluded. A partial remedy to these problems is provided 
by the Fierz-Pauli  framework in \cite{PF}, according to which the $\Phi$ tensor in (\ref{FP-tsr-rdct}) is taken as  
traceless and symmetric with respect to the $(\xi^i{\stackrel{\bullet}{\beta}_i})$ pairs of spinor indexes (each pair being equivalent to a Lorentz index) 
\begin{eqnarray}
\mbox{FP}:\quad \quad  j=n\in D^{\left( \frac{n}{2},\frac{n}{2}\right)}:\quad \mbox{Sym}\, \Phi^{\alpha_1}_{\stackrel{\bullet}{\beta}_1}{}^{...}_{...}{}^{\alpha_i}_{\stackrel{\bullet}{\beta}_i}{}^{...}_{...}{}^{\alpha_j}_{\stackrel{\bullet}{\beta}_j}{}^{...}_{...}{}
^{\alpha_n}_{\stackrel{\bullet}{\beta}_n} &=&\mbox{Sym}\, \Phi^{\alpha_1}_{\stackrel{\bullet}{\beta}_1}{}^{...}_{...}{}^{\alpha_j}_{\stackrel{\bullet}{\beta}_j}{}^{...}_{...}{}^{\alpha_i}_{\stackrel{\bullet}{\beta}_i}{}^{...}_{...}{}^{\alpha_n}_{\stackrel{\bullet}{\beta}_n},\label{Tnsr_FP}\\
\mbox{tr}\, \mbox{Sym}\, \Phi^{\alpha_1}_{\stackrel{\bullet}{\beta}_1}{}^{...}_{...}{}^{\alpha_j}_{\stackrel{\bullet}{\beta}_j}{}^{...}_{...}{}^{\alpha_i}_{\stackrel{\bullet}{\beta}_i}{}^{...}_{...}{}^{\alpha_n}_{\stackrel{\bullet}{\beta}_n}&=&0,
\label{FP_SymTrls}
\end{eqnarray}
and conditioned by,
\begin{eqnarray}
\partial_{\alpha_i}{}^{\stackrel{\bullet}{\beta_j}}\mbox{Sym}\, \Phi^{\alpha_1}_{\stackrel{\bullet}{\beta}_1}{}^{...}_{...}{}^{\alpha_j}_{\stackrel{\bullet}{\beta}_j}{}^{...}_{...}{}^{\alpha_i}_{\stackrel{\bullet}{\beta}_i}{}^{...}_{...}{}^{\alpha_n}_{\stackrel{\bullet}{\beta}_n}&=&0.
\label{Aux1}
\end{eqnarray}
The latter equation  acts as an auxiliary condition to the dynamics introduced by  setting the  $\Phi$  tensor-spinor on its mass-shell,
\begin{eqnarray}
(\partial^2+m^2)\mbox{Sym}\, {\Phi}^{\alpha_1}_{\stackrel{\bullet}{\beta_1}}{}^{...}_{...}{}{}^{\alpha_n}_{\stackrel{\bullet}{\beta_n}}\, &=&0.
\label{FierzPauli_Dotted_Undotted}
\end{eqnarray}
Though the FP approach \cite{PF} circumvents  some of the inconsistencies typical of high-order theories, such as the Ostrogradskian instability \cite{TaiJunChen}, it is still not completely consistent as it does not exclude ghosts and the propagation of its solutions upon coupling to an electromagnetic background can violate causality, this  basically because of the violation of the auxiliary condition
(\ref{Aux1}) in the presence of interaction.\\

\noindent
To recapitulate, the BW scheme has the advantage to amount to one single  wave equation, however of a high-order, while the FP method amounts to a second order equation but invokes an auxiliary condition, difficult to tackle upon gauging.
The goal of the present work is to combine the advantages of both approaches and avoid their difficulties.
{}For this purpose, we shall be seeking to 
\begin{itemize}
\item construct within the general tensor-spinor basis $S^r_s$ in (\ref{Gl_prm}) with $r+s=n$,
a  second order  differential equation invoking, as in the FP approach,  the on-mass-shell condition,  but in  such a way  that the equation is
$(j,0)\oplus (0,j)$ representation specific, free from inconsistencies  and,  differently from the Fierz-Pauli framework,  without  auxiliary conditions.
\end{itemize}
 Our case is that the  Lorentz group, when acting exclusively on the internal spin degrees of freedom,
does indeed provide the adequate  tools for the realization of such a program.
Below we shall see that in working in the whole space of the $S^r_s$ tensor-spinors in (\ref{Gl_prm}), dropping the conditions on the on-mass-shell-ness of the ``constituent'' spinors  
and the symmetry of the spinor indexes in (\ref{SYM}), the uncomfortable  high $\partial ^{2j}$ order of the related  differential wave equations of the BW scheme  becomes replaced by a second order equation but without any need for auxiliary conditions.

However, before turning to the next section we wish to note that the Bargmann-Wigner approach is supposed to be consistent with 
that developed by Joos and Weinberg in \cite{Joos}, \cite{Weinberg}, respectively, 
and in which  the single-spin-$j$  is described in terms of  a set of the even number of $2(2j+1)$ functions, constituting a so called ``bi-vector'', as
\begin{eqnarray} 
\psi^{(j)}=\left(\begin{array}{c}
\psi^{(j)}_1\\
...\\
\psi^{(j)}_{2j+1}\\
\psi^{(j)}_{(2j+1)+1}\\
...\\
\psi^{(j)}_{2(2j+1)}
\end{array}
\right)  &\simeq & (j,0)\oplus (0,j).
\label{WJ-spinors} 
\end{eqnarray}
The $2(2j+1)$-component wave function $\psi^{(j)}$ 
satisfies a differential matrix equation of the same high-order as the Bargmann-Wigner equation, given by,
\begin{eqnarray} 
\left(i^{2j}\left[ 
\gamma_{\mu_1\mu_2...\mu_{2j}}\right]_{AB}\partial^{\mu_1}\partial^{\mu_2}... 
\partial^{\mu_{2j}}-m^{2j}\delta_{AB}\right)\psi^{(j)}_B(x)&=&0, \quad B \in \left[1, 2(2j+1) \right].
\label{WJ-eqs} 
\end{eqnarray} 
Here, $\left[ \gamma_{\mu_1\mu_2...-\mu_{2j}}\right]_{AB}$ are 
the elements of the 
generalized Dirac Hermitian matrices of dimensionality 
$\left[2(2j+1)\right]\times \left[2(2j+1)\right]$, which 
transform as Lorentz tensors of rank-$2j$. 
The complete sets of such matrices have been extensively studied in the 
literature for the purpose of constructing all the possible  bi-linear forms of the field needed in the 
definitions of the 
generalized  currents, both  transitional  and diagonal \cite{Sankar}. Notice that the wave equation in (\ref{WJ-eqs}) emerges as a similarity transformation of the parity operator within the bi-vector space by the corresponding operator of the boost, which is of order $j$ in the momenta and encodes the frame dependence of the representation space. In consequence, the spin-$j$ degrees of freedom within the Joos-Weinberg method transform
according to strictly irreducible $(j,0)\oplus(0,j)$ representation spaces of the Lorentz group algebra. Below we shall bring an example which shows  that in reality
the Bargmann-Wigner method is not fully consistent with the Joos-Weinberg approach because it does not necessarily guarantee the irreducibility of its predicted  $2(2j+1)$ degrees of freedom under Lorentz transformations. We then suggest an upgrade of the BW equations by which the aforementioned inconsistency is removed. 

\subsection{The algebra of the spin-Lorentz group,  its Casimir invariants, and  momentum independent Lorentz projectors on 
irreducible representation spaces.}
The Lorentz group transforming the internal spin degrees of freedom, henceforth termed to as  spin- Lorentz group, and denoted by ${\mathcal L}$,  is a subgroup of the complete Lorentz group, which  acts besides on the spin- also  on the external space time.
The ${\mathcal L}$ generators, denoted by $S_{\mu\nu}$, are quadratic $d\times d$ constant matrices, where $d$ fixes the finite dimensionality of the internal representation space, and encodes the spin value. For the special case of a pure spin, dimensionality and spin are related as
$d=2(2j+1)$, while for representations of multiple spins, relations like  $d=\sum_i(2j_i+1)$, or, $ d=2\sum_i (2j_i+1)$, can hold valid.
Now, the direct product of ${\mathcal L}$ and  ${\mathcal T}_4$, the group of translations in the external space-time,
 whose generators, $i\partial _\mu$,  represent the quantum mechanical operators, $P_\mu$,  
of the components of the relativistic four-momentum, i.e. $P_\mu=i\partial_\mu$,  is generated by the following sub-set of the Poincar\'e algebra:
\begin{eqnarray}
{\mathcal L}:\quad \left[S_{\mu\nu},S_{\rho\sigma}\right] &=&i(g_{\mu\rho}S_{\nu \sigma}-g_{\nu\rho}S_{\mu\sigma}+g_{\mu\sigma}S_{\rho\nu}-g_{\nu\sigma}S_{\rho\mu}),
\label{Lrntz_algbr}\\
{\mathcal T}_4:\qquad \left[ P_\mu,P_\nu \right]&=&0,\label{trnsl_algbr}\\
\left[S_{\mu\nu},P_\lambda \right] &=&0,
\label{intrtwn}
\end{eqnarray}
to be termed to as the ``inhomogeneous spin-Lorentz group''.  In contrast, within the  algebra of the full  Poincar\'e group 
(also termed to as ``inhomogeneous Lorentz group'') the commutators in (\ref{intrtwn}) are non-vanishing because there,
the Lorentz group generators in the internal spin space  are supplemented
by the angular momentum and boost operators, $L_{\mu\nu}=-i\left(x_\mu \partial ^\nu  -x_\nu \partial^\mu \right)$,
which transform the external space time, and which do not commute with the operators of translations.
The  algebra of the spin-Lorentz group, given in (\ref{Lrntz_algbr}),  has two Casimir invariants  \cite{KimNoz},
here denoted in their turn by, $F$ and $G$, and  defined as,
\begin{eqnarray}
F_{AB}&=& \frac{1}{4}\left[S^{\mu\nu}\right]_{AD}\left[S_{\mu\nu}\right]_{DB}, \label{Cas1}\\
G_{AB}&=&\frac{1}{8}\epsilon_{\mu\nu\alpha\beta}\left[S^{\mu\nu}\right]_{AC}\left[S_{\alpha\beta}\right]_{CB}, \quad A,B, C, D,...=1,...,d.
\label{Cas2}
\end{eqnarray}
Due to the constancy of the quadratic $d\times d$ dimensional matrices $S_{\mu\nu}$, the  $F$, and $G$ operators commute both with the generators of translation, 
$P_\mu$, and with $P^2$, one of the Casimir invariants of the algebra of the full Poincar\'e group. 
In fact, the Casimir invariants of the algebra of the spin-Lorentz group, behave as invariants of 
the  algebra, ${\mathcal T}_4\times {\mathcal L}$, which generates the ``inhomogeneous spin-Lorentz group''. 
This is a remarkable property which we will put at work in what follows.
The two operators in (\ref{Cas1}) and (\ref{Cas2}), have the property of unambiguously identifying  any \underline{irreducible}
finite dimensional ${\mathcal L}$ group representation space, here generically denoted by, $D^{(j_1,j_2)\oplus (j_2,j_1)}$,  through their eigenvalues according to,
\begin{eqnarray}
F \, D^{(j_1,j_2)\oplus (j_2,j_1)}&=&c_{(j_1,j_2)}D^{ \left(j_1,j_2 \right)\oplus (j_2,j_1)},
\label{F_Cas}\\
c_{(j_1,j_2)}&=&\frac{1}{2}\left(K(K+2)+M^2\right), \quad K=j_1+j_2, \quad M=|j_1-j_2|,
\label{F_csts}\\
G\, \, D^{\left(j_1,j_2\right)}&=&r_{(j_1,j_2)}D^{\left(j_1,j_2\right)}, \quad G\, \, D^{(j_2,j_1)}=r_{(j_2,j_1)}D^{\left(j_2,j_1\right)}
\label{G_Cas}\\
r_{(j_1,j_2)}&=&-r_{(j_2,j_1)}=i(K+1)M.
\label{G_csts}
\end{eqnarray}
The idea of the present work, exposed in the next section, is to employ the Casimir invariant $F$ in the construction of a momentum independent (static) projector on the irreducible sectors of
the $n=2j$ rank spinor  in (\ref{Gl1}) and to explore the consequences. The article closes with brief conclusions and has one appendix, devoted to the Lagrangian description and the coupling to an electromagnetic field.\\

\noindent
\section{ Momentum independent projector on the $D^{(n/2,0)\oplus (0,n/2)}$ irreducible sector 
of the rank-$(2j)$  tensor-spinors. The spin-Lorentz group projector method.} 

We here are specifically interested in projectors on the irreducible Lorentz representations  appearing in the rhs of the equation (\ref{Gl1}) which contain  spin $j=n/2$.
The first  projector we wish to consider, here denoted by ${\mathcal P}^{(n/2,0)}$, is the one that identifies the $D^{(n/2,0)\oplus (0,n/2)}$ irrep.
We construct it from $F$ in (\ref{F_Cas}) as
\begin{equation}
\mathcal{P}_F^{(n/2,0)}= \Pi_{k l} \times \left( \dfrac{F-c_{(j_k,j_l)}}{c_{(n/2,0)}-c_{(j_k,j_l)}} \right), \label{General_L_Proj}
\end{equation}
where, $\Pi_{kl} \times$ denotes the ordinary product of the  operators in parenthesis, the pairs of indexes $(k,l)$ run over all the $D^{(j_k,j_l)}$ labels characterizing  the  irreducible
representation spaces in the rhs of (\ref{Gl1}), and the constants  $ c_{(j_k,j_l)}$
are those defined in (\ref{F_csts}). The equation (\ref{General_L_Proj})
shows that the operator $\mathcal{P}_F^{(n/2,0)}$ has the property to nullify any irreducible representation space for which
$(j_k,j_l)\not= (n/2,0)$. Instead, for $(j_k,j_l) = (n/2,0)$, it acts as the identity operator, meaning that $\mathcal{P}_F^{(n/2,0)}$
is a projector on $D^{(n/2,0)\oplus (0,n/2)}$.
It is obvious that $\mathcal{P}_F^{(n/2,0)}$ is of zeroth order in the momenta (the derivatives).
That is possible due to the direct product character, ${\mathcal T}_4\times {\mathcal L}$ of the symmetry group.
Indeed, the spin-Lorentz group ${\mathcal L}$, in factorizing from the group of translations, and in exclusively acting on the internal space,
does not need external transformations for the identification of its irreducible  degrees of freedom.
Furthermore, it is sufficient to require the $D^{(n/2,0)\oplus (0,n/2)}$ states to be on their mass-shell, and drop
the requirement of on-mass-shell-ness of the ``constituent'' Dirac spinors. The reward will be a 
$D^{(n/2,0)\oplus (0,n/2)}$-specific  second order differential equation within  the general $S^r_s$  spinor-tensor  basis in (\ref{Gl1}).
In the following we formulate the method of tracking down the $D^{(j,0)\oplus (0,j)}$ sector in (\ref{Gl1})   in terms of ${\mathcal P}_F^{(n/2,0)}$, and explore consequences. However, before moving to this issue, we shall switch from the spinor--co-spinor (chiral)--- to the commonly used  Dirac's  parity representation according to,
\begin{eqnarray}
\left( 
\begin{array}{c}
\xi^1\\
\xi^2\\
\eta_{\stackrel{\bullet}{1}}\\
\eta_{\stackrel{\bullet}{2}}
\end{array}
\right)\longrightarrow
\left( 
\begin{array}{c}
\xi^1+\eta_{\stackrel{\bullet}{1}}\\
\xi^2+\eta_{\stackrel{\bullet}{2}}\\
\left(\eta_{\stackrel{\bullet}{1}}-\xi^1\right)\\
\left(\eta_{\stackrel{\bullet}{2}}-\xi^2\right)
\end{array}
\right)\equiv\left(
\begin{array}{c}
\psi_1\\
\psi_2\\
\psi_3\\
\psi_4
\end{array}
\right),\label{part_rpr}
\end{eqnarray}
in which case the dotted and undotted indexes are all replaced by regular Dirac indexes, $a_i=1,2,3,4$, for which we shall use small Latin letters. In what follows we systematically use the conventional Dirac spinors  in the composition of the  spinor- tensors meaning that instead of $\left( S^{r}_s\right){}^{\alpha_1...\alpha_r}{}_{\stackrel{\bullet}{\beta_1}...\stackrel{\bullet}{\beta_s}}$ in (\ref{Gl_prm}), with $r+s=n$,
we shall be using Dirac-spinor--tensors of rank-$n$, here denoted by $\Psi^{(n)}{}_{a_1a_2... a_n}$ with $a_i=1,2,3,4$, i.e.
the following notational change will be undertaken,
\begin{equation}
\left( S^{r}_s\right){}^{\alpha_1...\alpha_r}{}_{\stackrel{\bullet}{\beta_1}...\stackrel{\bullet}{\beta_s}}\longrightarrow
\Psi^{(n)}{}_{a_1a_2... a_n}, \quad a_i=1,2,3,4, \quad r+s=n.
\label{our_basis}
\end{equation}

\noindent
\subsection{One sole second order differential wave equation for spin-$j\geq 1$ in the basis of the tensor-spinor of rank-$2j$ .}
The  Lorentz group generators, $\left[ S^{(1)} \right]_{\mu \nu}$,  in  the Dirac spinor space, $D^{(1/2,0)\oplus(0,1/2)}$, are textbook knowledge \cite{KimNoz} and read,
\begin{equation}
\left[ S^{(1)} \right]_{\mu \nu}=\dfrac{i}{4}[\gamma_\mu, \gamma_\nu]=\dfrac{1}{2} \sigma_{\mu \nu}, \quad \mu, \nu=0,1,2,3,
\label{Spinor_generators}
\end{equation}
where $\gamma_{\mu}$, and $\gamma_\nu$ stand for the standard Dirac matrices. Then,  the spin-Lorentz group generators, $\left[ S^{(n)} \right]_{\mu \nu}$,  
in the reducible  $\Psi^{(n)}_{a_1a_2...a_n}=\otimes _{i=1}^{i=n} D^{\left[(1/2,0)\oplus (0,1/2)\right]_i}$ basis in (\ref{Gl_prm}), and  in the new notations of (\ref{our_basis}),  are calculated following the standard prescription regarding generator construction in product spaces as,
\begin{equation}
\left[ S^{(n)} \right]_{\mu \nu}=\left[ S^{(1)} \right]_{\mu \nu} \otimes \left[ \Pi_{i=1}^{i=n-1} \otimes \mathbf{1}\right]
+ \mathbf{1} \otimes \left[ S^{(1)} \right]_{\mu \nu}\left[  \otimes \Pi_{i=1}^{i=n-2} \otimes \mathbf{1}\right]+ \cdots +
\left[ \Pi_{i=1}^{i=n-1} \otimes \mathbf{1}\right] \otimes \left[ S^{(1)} \right]_{\mu \nu},
\label{Generator_direct_product}
\end{equation}
with $\left[ S^{(1)} \right]_{\mu \nu}$ in (\ref{Spinor_generators}), and  $\mathbf{1}$ standing for the $4 \times 4$ unit matrix in the Dirac spinor space.  In substituting  (\ref{Spinor_generators}) into (\ref{Generator_direct_product}), and then (\ref{Generator_direct_product}) into (\ref{General_L_Proj}),
the $P_F^{(n/2,0)}$ projector of interest is explicitly constructed. Once again, it is seen that this  projector is a static one and does not provide a wave equation. In order to introduce the free kinematics, we impose on the states spanning the $D^{(n/2,0)\oplus (0,n/2)}$ representation space the indispensable mass-shell condition via the Klein-Gordon equation, and find the following master equation,
\begin{equation}
\mbox{This}\,\, \mbox{work:}\quad \left( \partial_\mu \partial^\mu \left[\mathcal{P}^{(n/2,0)}_{F}\right]_{a_1 a_2\dots a_{n}}^{~~~~~~~~~~~~~~~ b_1 b_2 \dots b_{n}}   + m^2 \right) \Psi^{(n)}_{b_1 b_2 \dots b_{n}}=0, \quad n=2j. \label{second_order_equation_spinorialtensor}
\end{equation}
\begin{quote}
This is the one sole  second order differential equation which we suggest in place of (i) the high-order Bargmann-Wigner coupled system of equations in (\ref{BW-1})-(\ref{BW-2}), and (ii)  the Fierz-Pauli system of equations (\ref{Aux1})-(\ref{FierzPauli_Dotted_Undotted}).
\end{quote}  
As long  as within the Bargmann-Wigner  framework in (\ref{BW-1})-(\ref{BW-2}), the two  $D^{(n/2.0)}$, and  $D^{(0,n/2)}$
building blocks of the $2(2j+1)$ spin-$j=n/2$ degrees of freedom are not searched at all by the
wave operator but are put by hand through the symmetrization of the tensor indexes, they transform irreducibly under $SU(2)$ but, as we shall see below, their sum is reducible under Lorentz transformations. Instead,  the wave operator in our suggested master equation  (\ref{second_order_equation_spinorialtensor}), in   acting on the complete set of  indexes characterizing the Lorentz reducible tensor-spinors,  unambiguously identifies in it  the  Lorentz \underline{irreducible} $D^{(n/2,0)\oplus (0,n/2)}$  sector, while neatly cutting out the rest. 
The Fierz-Pauli method  identifies the spin-$j$ degrees of freedom implicitly and by the aid of the auxiliary conditions,  which is not conserved by the gauging procedure, and needs special adjustments. 
Stated differently, the master  equation (\ref{second_order_equation_spinorialtensor}) describes a manifestly genuine and strictly representation specific eigenvalue problem that is free  from auxiliary conditions.
It  identifies the spin-$j$ degrees of freedom  directly, at once, and unambiguously, a reason for which it can be derived from a Lagrangian, and 
 coupled to the electromagnetic field in the regular way of a minimal gauging (see the Appendix).

In this manner, we furnished one single  second order differential wave equation for a particle of spin-$j=n/2$ which we described
in terms of  a general (Dirac)spinor--tensor  of rank $2j$. In the following, we work out for illustrative purposes  
the case of spin-$1$ residing in the $D^{(1,0)\oplus (0,1)}$ irreducible sector of second-rank (Dirac)spinor-tensor, $\Psi^{(2)}$.\\

\noindent
\subsection{The wave equation of spin-$1$ in $\Psi^{(2)}$. An illustrative example.}
As an illustrative example for our suggested method we consider the simplest case of an integer spin, namely,  spin-$1$ as embedded by 
tensor-spinors of second rank,
$\Psi^{(2)}_{a_1 a_2}=\psi_{a_1}\psi_{a_2}\simeq D^{(1/2,0)\oplus (0,1/2)}\otimes  D^{(1/2,0)\oplus (0,1/2)}$, with $\psi_{a_i}$ defined in (\ref{our_basis}).
The latter tensor is reducible under Lorentz transformations according to,
\begin{equation}
\Psi^{(2)}_{a_1a_2}\simeq D^{(1/2,0)\oplus(0,1/2)}\otimes D^{(1/2,0)\oplus(0,1/2)}=D^{(1,0)\oplus(0,1)}\oplus 2D^{(0,0)} \oplus 2D^{(1/2,1/2)},
\label{spin1_1}
\end{equation}
where the integer numbers in front of the irreps stand for their multiplicities in the direct sum.
The rhs in (\ref{spin1_1}) contains three different irreducible Lorentz representation spaces, whose eigenvalues with respect to the Casimir invariant $F$ in (\ref{F_Cas}) are calculated from the expressions given in (\ref{F_csts}) as,
\begin{eqnarray}
c_{(0,0)} &=&0 \quad \quad \text{for} \quad (0,0), \\
c_{(1,0)} &=&2 \quad \quad \text{for} \quad (1,0)\oplus(0,1), \\
c_{(1/2,1/2)} &=&\frac{3}{2} \quad \quad \text{for} \quad (1/2,1/2).
\end{eqnarray}
Therefore, according to (\ref{General_L_Proj}), the projector ${\mathcal P}_F^{(1,0)}$ on $(1,0)\oplus (0,1)$  emerges as,
\begin{equation}
\mathcal{P}_F^{(1,0)}=\dfrac{1}{2}\left( 2F^2-3F \right). \label{Lor_proj_form1}
\end{equation}
Next we evaluate  the generators within $\Psi^{(2)}_{a_1 a_2}$  by the aid of (\ref{Generator_direct_product}) setting  $n=2$,
substitute them  in (\ref{Cas1}), and then insert the result for $F$ in   (\ref{General_L_Proj}).
In so doing we calculate the following explicit expression for the $F$- Casimir invariant of the spin-Lorentz group algebra:
\begin{eqnarray}
\Psi_{a_1a_2}^{(2)}:\quad F&=&\dfrac{1}{16} \left( \mathbf{1} \otimes \sigma^{\mu \nu} + \sigma^{\mu \nu} \otimes \mathbf{1} \right)
\left( \mathbf{1} \otimes \sigma_{\mu \nu} + \sigma_{\mu \nu} \otimes \mathbf{1} \right).
\label{spin1_2}
\end{eqnarray}
In index notation,  $F$ reads,
\begin{equation}
F^{a_1 a_2}_{~~~~~~~~b_1 b_2} =\dfrac{1}{8}\left[ 12\; \delta^{a_1}_{~~~~b_1} \delta^{a_2}_{~~~~b_2} + \left( \sigma_{\mu \nu} \right)^{a_1}_{~~~~b_1} \left( \sigma^{\mu \nu} \right)^{a_2}_{~~~~b_2} \right]. \label{F_spinor_components}
\end{equation}
Substitution of (\ref{F_spinor_components}) into (\ref{Lor_proj_form1}) amounts to the following explicit expression for the
searched spin-$1$ Lorentz projector:
\begin{eqnarray}
\left[ \mathcal{P}_F^{(1,0)} \right]^{a_1 a_2}_{~~~~~~~~b_1 b_2}&=&\dfrac{1}{32}
\left( \sigma_{\mu \nu} \right)^{a_1}_{~~~~b'_1}\left( \sigma_{\mu \nu} \right)^{a_2}_{~~~~b'_2} \left[ 12\; \delta^{b'_1}_{~~~~b_1} \delta^{b'_2}_{~~~~b_2} + \left( \sigma_{\mu \nu} \right)^{b'_1}_{~~~~b_1} \left( \sigma^{\mu \nu} \right)^{b'_2}_{~~~~b_2} \right] \label{Lor_proj_form_components}  \\ \nonumber  
&=& \dfrac{1}{4}
\left( \sigma_{\mu \nu} \right)^{a_1}_{~~~~b'_1}\left( \sigma_{\mu \nu} \right)^{a_2}_{~~~~b'_2}
F^{b'_1 b'_2}_{~~~~~~~~b_1 b_2}.
\label{our_spin1_projector}
\end{eqnarray}
Therefore, according to (\ref{second_order_equation_spinorialtensor}), the second order differential  wave equation for a particle of spin-$1$
described in terms of a second rank Dirac spinor takes the following form:
\begin{equation}
\left( \partial_\mu \partial^\mu 
\left[ \mathcal{P}_F^{(1,0)} \right]_{a_1 a_2}^{~~~~~~~~b_1 b_2}+ m^2 \right)
\Psi^{(2)}_{b_1 b_2}=0. \label{Wave_equation_spin1}
\end{equation}
Along this line, any arbitrary spin-$j$ can be described by means of a rank-$2j$ (Dirac)spinor-tensor satisfying a second order differential equation.
Below, the solutions to the equation (\ref{Wave_equation_spin1}), when considered in the momentum space upon the replacement, $i\partial_\mu \to p_\mu$,  are obtained, and compared to the corresponding  solutions appearing in the  Bargmann-Wigner framework.\\

\subsection{ Comparison of spin-$1$  predictions following from the respective spin-Lorentz group projection method,  and  the
 Bargmann-Wigner framework}
A set of linearly independent solutions in  momentum space for the spin-$1$ states satisfying  the equation in (\ref{Wave_equation_spin1})
can be constructed upon applying the Lorentz projector $\mathcal{P}_F^{(1,0)}$ in (\ref{Lor_proj_form_components}) to the sixteen dimensional
rank-2 tensor-spinors, $\left[ u_\pm(\mathbf{p},\lambda) \right]^a \left[ u_\pm(\mathbf{p},\lambda) \right]^b$, as composed by the momentum space
Dirac spinors of positive ($+$),  and negative, ($-$), parities,
\begin{eqnarray}
u_{+}(\mathbf{p},1/2)\equiv  u(\mathbf{p},1/2)&=&\dfrac{1}{\sqrt{2m(m+p_0)}}\begin{pmatrix}
m+p_0 \\ 0 \\ p_3\\ p_1+ip_2
\end{pmatrix}, \nonumber\\\
u+(\mathbf{p},-1/2)\equiv u(\mathbf{p},-1/2)&=&\dfrac{1}{\sqrt{2m(m+p_0)}}\begin{pmatrix}
0\\ m+p_0\\ p_1-ip_2 \\ -p_3
\end{pmatrix}, \nonumber\\
u_{-}(\mathbf{p},1/2)\equiv v(\mathbf{p},1/2)&=&\dfrac{1}{\sqrt{2m(m+p_0)}} \begin{pmatrix}
p_3\\p_1+ip_2\\m+p_0\\0
\end{pmatrix},\nonumber\\
u_{-}(\mathbf{p},-1/2)\equiv v(\mathbf{p},-1/2)&=&\dfrac{1}{\sqrt{2m(m+p_0)}} \begin{pmatrix}
p_1-ip_2\\-p_3\\0\\m+p_0
\end{pmatrix}.
\end{eqnarray}
The sixteen dimensional rank-$2$ Dirac spinors  span the  reducible Lorentz representation space $D^{(1/2,0) \oplus (0,1/2)} \otimes D^{(1/2,0) \oplus (0,1/2)}$. Executing the proper calculation, we find precisely  six linearly independent combinations, as it should be, and  list them in the Table \ref{Comparative_table_basis} together with the corresponding Bargmann-Wigner spin-$1$ states, as a comparison.

\begin{table}[h]
\begin{tabular}{|c | c|}
\hline
Spin-$1$  states (this work) & Spin-$1$ Bargmann-Wigner states \\ [0.5ex] \hline

$\begin{array}{l c l}
\left[ w^{(2)}_+(\mathbf{p},1) \right]^{a b}&=&\frac{1}{\sqrt{2}} \big( \left[ u_+(\mathbf{p},1/2) \right]^a \left[ u_+(\mathbf{p},1/2) \right]^b \\
& & + \left[ u_-(\mathbf{p},1/2) \right]^a \left[ u_-(\mathbf{p},1/2) \right]^b \big)
\end{array}$ &
$\left[ W_+^{(2)}(\mathbf{p},1) \right]^{a b}= \left[ u_+(\mathbf{p},1/2) \right]^a \left[ u_+(\mathbf{p},1/2) \right]^b$ \\ [0.5cm] \hline

$\begin{array}{l c l}
\left[ w^{(2)}_+(\mathbf{p},0) \right]^{ab}&=&\frac{1}{2} \big( \left[ u_+(\mathbf{p},1/2) \right]^a\left[ u_+(\mathbf{p},-1/2) \right]^b \\
& & + \left[ u_+(\mathbf{p},-1/2) \right]^a \left[ u_+(\mathbf{p},1/2) \right]^b \\
& & + \left[ u_-(\mathbf{p},1/2) \right]^a \left[ u_-(\mathbf{p},-1/2) \right]^b \\
& & + \left[ u_-(\mathbf{p},-1/2) \right]^a \left[ u_-(\mathbf{p},1/2) \right]^b \big)
\end{array}$ &
$\begin{array}{l c l}
\left[ W_+^{(2)}(\mathbf{p},0) \right]^{a b}&=& \left[ u_+(\mathbf{p},1/2) \right]^a \left[ u_+(\mathbf{p},-1/2) \right]^b \\
& & + \left[ u_+(\mathbf{p},-1/2) \right]^a \left[ u_+(\mathbf{p},1/2) \right]^b
\end{array}$\\ [1cm] \hline

$\begin{array}{l c l}
\left[ w^{(2)}_+(\mathbf{p},-1) \right]^{ab}&=&\frac{1}{\sqrt{2}} \big( \left[ u_+(\mathbf{p},-1/2) \right]^a \left[ u_+(\mathbf{p},-1/2) \right]^b \\
& & + \left[ u_-(\mathbf{p},-1/2) \right]^a\left[ u_-(\mathbf{p},-1/2) \right]^b \big)
\end{array}$ &
$\left[ W_+^{(2)}(\mathbf{p},-1) \right]^{a b}= \left[ u_+(\mathbf{p},-1/2) \right]^a \left[ u_+(\mathbf{p},-1/2) \right]^b$ \\ [0.5cm] \hline

$\begin{array}{l c l}
\left[ w_-^{(2)}(\mathbf{p},1) \right]^{ab}&=&\frac{1}{\sqrt{2}} \big( \left[ u_+(\mathbf{p},1/2) \right]^a \left[ u_-(\mathbf{p},1/2) \right]^b\\
& & + \left[ u_-(\mathbf{p},1/2) \right]^a \left[ u_+(\mathbf{p},1/2) \right]^b \big)
\end{array}$ &
$\left[ {\widetilde W}_+^{(2)}(\mathbf{p},1) \right]^{ab }= \left[ u_-(\mathbf{p},1/2) \right]^a \left[ u_-(\mathbf{p},1/2) \right]^b$ \\ [0.5cm] \hline

$\begin{array}{l c l}
\left[ w^{(2)}_-(\mathbf{p},0) \right]^{ab}&=&\frac{1}{2} \big( \left[ u_+(\mathbf{p},1/2) \right]^a \left[ u_-(\mathbf{p},-1/2) \right]^b \\
& & + \left[ u_-(\mathbf{p},-1/2) \right]^a \left[ u_+(\mathbf{p},1/2) \right]^b \\
& & + \left[ u_+(\mathbf{p},1/2) \right]^a \left[ u_-(\mathbf{p},-1/2) \right]^b \\
& & + \left[ u_-(\mathbf{p},-1/2) \right]^a \left[ u_+(\mathbf{p},1/2) \right]^b \big)
\end{array}$ &
$\begin{array}{l c l}
\left[ {\widetilde W}_+^{(2)}(\mathbf{p},0) \right]^{a b}&=& \left[ u_-(\mathbf{p},1/2) \right]^a \left[ u_-(\mathbf{p},-1/2) \right]^b \\
& & + \left[ u_-(\mathbf{p},-1/2) \right]^a \left[ u_-(\mathbf{p},1/2) \right]^b
\end{array}$\\ [1cm] \hline

$\begin{array}{l c l}
\left[ w_-^{(2)}(\mathbf{p},-1) \right]^{ab}&=&\frac{1}{\sqrt{2}} \big( \left[ u_+(\mathbf{p},-1/2) \right]^a \left[ u_-(\mathbf{p},-1/2) \right]^b \\
& & + \left[ u_-(\mathbf{p},-1/2) \right]^a \left[ u_+(\mathbf{p},-1/2) \right]^b\big)
\end{array}$ &
$\left[ {\widetilde W}_+^{(2)}(\mathbf{p},-1) \right]^{a b}= \left[ u_-(\mathbf{p},-1/2) \right]^a \left[ u_-(\mathbf{p},-1/2) \right]^b $ \\ [0.5cm] \hline
\end{tabular}
\caption{Comparison  between the spin-$1$ states obtained in this work (left column) with the spin-$1$ states predicted by the Bargmann-Wigner equations (right column).}
\label{Comparative_table_basis}
\end{table}

\noindent
In order to make the comparison between the two schemes manifest, we write down in the subsequent two equations the explicit rank-$2$ Dirac spinors following from the present work, $\left[w_\pm ^{(2)}({\mathbf p},\lambda) \right]^{a b}$, on the one side, and from the Bargmann-Wigner approach,
$\left[W_+ ^{(2)}({\mathbf p},\lambda) \right]^{a b}$-$\left[{\widetilde W}_+ ^{(2)}({\mathbf p},\lambda) \right]^{a b}$, on the other, focusing on the particular case of $\left[w_{+}^{(2)}({\mathbf p},1)\right]^{ab}$, and $\left[W_{+}^{(2)}({\mathbf p},1)\right]^{ab}$, for the sake of concreteness. In so doing, we find the following normalized states,

\begin{eqnarray}
\left[w^{(2)}_{+}({\mathbf p},1)\right]^{ab}=\frac{\sqrt{2}}{4m(m+p_0)}&&\nonumber\\
\left(
\begin{array}{cccc}
(m+p_0)^2 +p_3^2 & (p_1+ip_2)p_3 & 2p_3(m+p_0)&(p_1+ip_2)(m+p_0)\\
(p_1+ip_2)p_3& (p_1+ip_2)^2 &(p_1+ip_2)(m+p_0)&0\\
2p_3 (m+p_0)&(p_1+ip_2)(m+p_0)&(m+p_0)^2 +p_3^2 & (p_1+ip_2)p_3\\
(p_1+ip_2)(m+p_0)&0&(p_1+ip_2)p_3 &(p_1+ip_2)^2
\end{array}
\right),&&
\label{we}
\end{eqnarray}
and

\begin{eqnarray}
\left[W_{+}{}^{(2)}({\mathbf p},1)\right]^{ab}&=&\frac{1}{2m(m+p_0)}\left(
\begin{array}{cccc}
(m+p_0)^2& 0&p_3(m+p_0)&(m+p_0)(p_1+ip_2)\\
0&0&0&0\\
p_3(m+p_0)&0&p_0^2&p_3(p_1+ip_2)\\
(p_1+ip_2)(m+p_0)&0&p_3(p_1+ip_2)&(p_1+ip_2)^2
\end{array}\right).\nonumber\\
\label{they}
\end{eqnarray}
An inspection of these equations shows that the tensor calculated in the equation (\ref{we}) of the present work  has the following six independent degrees of freedom:
\begin{eqnarray}
w^{11}=w^{33}, &\quad&  w^{12}=w^{21}=w^{34}=w^{43}, \quad w^{13}=w^{31}, \nonumber\\
w^{22}=w^{44}, &\quad& w^{14}=w^{41}=w^{23}=w^{32}, \quad w^{24}=w^{42},
\label{dof_we}
\end{eqnarray}
and so does the Bargmann-Wigner solution in (\ref{they}), though theirs are distinct from ours. The six independent spin-$1$ degrees of freedom  
following from the Bargmann-Wigner framework are,
\begin{eqnarray}
W^{11}, &\quad& W^{13}=W^{31}, \quad W^{14}=W^{41}, \nonumber\\
W{}^{33}, &\quad& W^{34}=W^{43}, \quad W^{44},
\label{dof_they}
\end{eqnarray}
while the rest of the components are identically vanishing.
In addition, upon comparing the tensor-spinor in (\ref{we}) with the one in (\ref{they}), we observe  differences (modulo normalization factors), between the following four components,
\begin{equation}
w^{11}\not=W^{11}, \quad w^{22}\not=W^{22}, \quad w^{33}\not=W^{33}, \quad w^{12}\not=W^{12},
\end{equation}  
while the remaining components are equal. Below these differences are attributed to the reducibility of the BW states under Lorentz transformations.
In the last two equations we use ``rationalized'' notations in which we dropped  subscripts  and arguments in 
$\left[ w_{+}^{(2)}({\mathbf p },1)\right]^{ab}$--
$\left[ W^{(2)}_+({\mathbf p})\right]^{ab}$ for the sake of keeping the formulas  possibly  more transparent.\\

\noindent
Before proceeding further we  notice that the six degrees of freedom of the $D^{(1,0)\oplus (0,1)}$ representation space can equivalently be described in terms of the (also six) independent components of the totally anti-symmetric Lorentz tensor of second rank, 
$B^{\mu\nu}$,  according to
\begin{equation}
D^{(1,0)\oplus (0,1)}\simeq \left(
\begin{array}{cccc}
0&B^{01}&B^{02}& B^{03}\\
-B^{01}&0&B^{12}&B^{13}\\
-B^{02}&-B^{12}&0&B^{23}\\
-B^{03}&-B^{13}&-B^{23}&0
\end{array}
\right).
\label{inst1}
\end{equation}
In so doing,  spin-$1$ is described in terms of degrees of freedom equipped with Lorentz indexes. The link to the spinor-index notation  
considered in the present work is established by the following relation, 
\begin{eqnarray}
\left(
\begin{array}{cc}
f^1_1&f^1_2\\
f^2_1&f^2_2
\end{array}
\right)=
\left(
\begin{array}{cc}
(E^3-iH ^3)&(E^1-iE^2)-i(H^1-iH^2)\\
(E^1-iE^2)+i(H^1-iH^2 )&-(E^3-iH^3)
\end{array}
\right),
\label{inst2}
\end{eqnarray}
where, $f^\alpha_\beta$ is a spinor-tensor of second rank in the undotted indexes, while the components of the vector ${\mathbf E}$, and the axial vector ${\mathbf H}$ fields relate to  Lorentz indexes according to,
\begin{equation}
E^i=B^{0i}, \quad H^1=B^{23},\quad -H^2=B^{13}, \quad  H^3=B^{12}, \quad i=1,2,3.
\label{EH_Fmunu_map}
\end{equation}
Obviously, a basis choice  does not affect the calculated values of the physical observables.
Therefore, embedding  $(1,0)\oplus (0,1)$  in particular, and $(j,0)\oplus (0,j)$ in general  by bases distinct from the tensor-spinors, is a matter of mere convenience 
and does not affect the physical properties of the particles. In view of this, a spin-$1$ particle transforming in $(1,0)\oplus (0,1)$ has to be characterized by the same set of physical observables in all approaches to high-spins. 
In the following we present the case where the Bargmann-Wigner spin-$1$ spinors transform reducibly under Lorentz transformations, while our spinors, in parallel of those of Joos-Weinberg, transform strictly irreducibly, as it should be, if the spin under consideration were to reside entirely in $D^{(1,0)\oplus (0,1)}$. {}For this purpose we construct the Lorentz projector, ${\mathcal P}_F^{(1/2,1/2)}$, on  the $D^{(1/2,1/2)}$ irreducible sector in the rhs of (\ref{Gl1})
and let it act on the Bargmann-Wigner spin-$1$ spinors.

\noindent
\subsection{Lorentz reducibility of the spin-$1$ Bargmann-Wigner states.}
Without entering into technical details, we limit ourselves to report  that the projector of our interest is obtained along the line of the
reasoning presented above and its general form is found as,
\begin{equation}
\mathcal{P}_F^{(1/2,1/2)}=\dfrac{4}{3} ( 2F-F^2 ).
\label{4vprjct}
\end{equation}
In spinor index notation we calculated it as,
\begin{equation}
\left[ \mathcal{P}_F^{(1/2,1/2)} \right]^{a_1 a_2}_{~~~~~~~ b_1 b_2} = \delta^{a_1}_{~~~~ b_1} \delta^{a_2}_{~~~~ b_2}
-\dfrac{1}{12} \left( \sigma_{\mu_1 \nu_1}  \right)^{a_1}_{~~~~ c} \left( \sigma_{\mu_2 \nu_2}  \right)^{c}_{~~~~ b_1}
\left( \sigma^{\mu_1 \nu_1}  \right)^{a_2}_{~~~~ d} \left( \sigma^{\mu_2 \nu_2}  \right)^{d}_{~~~~ b_2}-
\dfrac{2}{12} \left( \sigma_{\mu \nu} \right)^{a_1}_{~~~~ b_1} \left( \sigma^{\mu \nu} \right)^{a_2}_{~~~~ b_2}.
\label{4v_spnindx}
\end{equation}
When applied to any of the Bargmann-Wigner  $\left[ W_+(\mathbf{p},\lambda) \right]^{ab}{}$-
$\left[{\widetilde W}_+ ({\mathbf p},\lambda) \right]^{ab }$ spinors, non-vanishing  projections on
$D^{(1/2,1/2)}$ states  (not shown here) are obtained. This means that the spin-$1$ states following from the  Bargmann-Wigner framework
are  linear combinations of spin-$1$ states residing in  the two distinct Lorentz irreducible representation spaces, $D^{(1,0)\oplus(0,1)}$, and
$D^{(1/2,1/2)}$. 
In other words, differently from  the supposed consistency with the Joos-Weinberg approach (discussed at the end of the section 1.1), and with our spin-$1$ degrees of freedom, the junction of the two  spin-$1$ triplets generated by the Bargmann-Wigner approach violates the Lorentz-irreducibility, a shortcoming of serious consequences.
Indeed, in order to satisfy Wigner's definition of an elementary particle at the classical level, its spin-degrees of freedom have to transform according 
to non-unitary finite dimensional irreducible representations of the spin-Lorentz group so that upon quantization, the states of continuous four-momenta could transform according to the infinite dimensional unitary representations of the Poincar\'e algebra. Within this context, the mixture of  irreducible spin-Lorentz group representations has to be considered as unphysical and removed. The reason behind the request on irreducibility is that particles of equal spins, transforming in distinct spin-Lorentz group representation spaces, can have different  physical properties.
{}For example, in ref.~\cite{PRD85_2012} the electromagnetic multipole moments of particles with spin-$1$ residing in the four-vector, $(1/2,1/2)$, on the one side, and within the strictly irreducible Joos-Weinberg spin-$1$ bi-vector $(1,0)\oplus (0,1)$
(equivalent to our approach according to the line of reasoning around (\ref{inst1})--(\ref{EH_Fmunu_map})), on the other side, have been explicitly calculated and compared. There, it was found that the Breit-frame  electric quadrupole $(E2)$ moments  of the two spin-$1$ particles under discussion  are of equal magnitudes, but of different orientations, amounting to opposite signs. Mixing the two representation spaces will therefore distort the $E2$ value. For this reason, compared to spin-1 be it in the spin-Lorentz group projection method, or the Joos-Weinberg method,  the BW spin-$1$ field  will be characterized by a different  electric quadrupole moment, unless its $(1/2,1/2)$ component has not been eliminated by the help of the Lorentz projector from (\ref{Lor_proj_form_components}). \\

\noindent
The $(1/2,1/2)$ component of the spin-$1$ BW field is easily eliminated upon application of the projector operator in (\ref{General_L_Proj}), (\ref{Lor_proj_form_components}) to
the eqs.~(\ref{BW-1})-(\ref{BW-2}) according to
\begin{eqnarray}
\left( \partial^{\alpha_1\stackrel{\bullet}{\beta_1}}\otimes \dots \otimes  \partial^{\alpha_n\stackrel{\bullet}{\beta_n}}\right) 
\left[ {\mathcal P}^{(j,0)}_F\right]
_{\stackrel{\bullet}{\beta_1}\dots\stackrel{\bullet}{\beta_n}}\,\,^{\stackrel{\bullet}{\tau_1}\dots\stackrel{\bullet}{\tau_n}}\mbox{Sym}{}\, \psi_{\stackrel{\bullet}{\tau_1}\dots\stackrel{\bullet}{\tau_n}} &=& (-im)^{n}\left[ {\mathcal P}^{(j,0)}_F\right]
_{\kappa_1\dots\kappa_n}\,\,^{\alpha_1\dots\alpha_n}\mbox{Sym}{}\, \psi{}^{\kappa_1\dots\kappa_n},
\label{BW-1P}\nonumber\\
\left(\partial_{\alpha_1\stackrel{\bullet}{\beta_1}}\otimes \dots \otimes \partial_{\alpha_n\stackrel{\bullet}{\beta_n}}\right)\left[{\mathcal P}^{(j.0)}_F\right]
_{\kappa_1\dots\kappa_n}\,\,^{\alpha_1\dots\alpha_n}
\mbox{Sym}{}\, \psi^{\kappa_1 \dots\kappa_n}&=&(-im)^n\left[ {\mathcal P}^{(j,0)}_F\right]
_{\stackrel{\bullet}{\beta_1}\dots\stackrel{\bullet}{\beta_n}}\,\,^{\stackrel{\bullet}{\tau_1}\dots\stackrel{\bullet}{\tau_n}} \mbox{Sym}{}\, 
\psi_{\stackrel{\bullet}{\tau_1}\dots\stackrel{\bullet}{\tau_n}},
\label{BW-2P}\nonumber\\
\end{eqnarray}
and setting $j=1$.
Upon this upgrade, the equivalence  between the tensor-spinors following from the Bargmann-Wigner-- and our spin-Lorentz group projector methods is achieved, and the consistency with the Joos-Weinberg approach is reached.
With that, the congruency among the predictions resulting from the three approaches under discussion regarding the physical observables characterizing the spin-$j$ particles transforming in $(j,0)\oplus (0,j)$
is warranted.

\noindent
In the Appendix we minimally gauge the equation in (\ref{second_order_equation_spinorialtensor}) and  prove its causality, besides presenting the corresponding Lagrangian.

\section{Conclusions.} In the present work we suggested an approach to the description of  any  spin-$j$ at the classical field theoretical level and by means of the one sole second order differential equation (\ref{second_order_equation_spinorialtensor}) operating on  the general rank-$2j$ (Dirac)spinor-tensor,
$\Psi^{(n)}$, defined  in (\ref{Gl1}).
The key ingredient in (\ref{second_order_equation_spinorialtensor}) is a momentum independent projector, 
$\left[{\mathcal P}_F^{(n/2,0)} \right]$, which we built up in (\ref{General_L_Proj}) and (\ref{Generator_direct_product}), and by the aid of 
(\ref{Spinor_generators}) from the Casimir invariant, $F$, in (\ref{F_Cas}) of the spin-Lorentz group,  
 according to the algebra in (\ref{Lrntz_algbr}) and (\ref{intrtwn}). This projector has the property to unambiguously identify
at once and as a whole the irreducible $D^{(j,0)\oplus (0,j)}$ sector of $\Psi^{(n)}$, while neatly 
sorting out the rest. 
The second order equation (\ref{second_order_equation_spinorialtensor}) is the one which replaces the high-order coupled equations in (\ref{BW-1}) and (\ref{BW-2}), on the one side, and the Fierz-Pauli second order framework in (\ref{Aux1}) and (\ref{FierzPauli_Dotted_Undotted}), on the other.  In being free from axillary conditions, the  equation 
(\ref{second_order_equation_spinorialtensor}) allows for a minimal gauging, is local and causal, thus avoiding the most serious problems
of Bargmann-Wigner's  higher-order equations and of the second order framework by Fierz and Pauli. In being based on the eigenvalue problem of a quadratic invariant form of the relativistic space time algebra, the suggested method  describes a pair of spin-$j$ particles of opposite parities.  
Fixing the parity of fermions would require further studies of the conditions for the bi-linearizations of the gauged second order wave equation. 
{}For example, it is well known  that the bi-linearization upon gauging  of the most general Klein-Gordon operator,
\begin{equation}
\partial ^\mu \partial_\mu -g\frac{i\sigma^{\eta\rho}}{4}\left[ \partial_\eta ,\partial_\rho\right]+m^2,
\end{equation}
 to the gauged Dirac equation (and its conjugate) is possible only for $g=2$ \cite{Greiner}.
However, for bosons, the fields of main interest here,  where particles and anti-particles are of equal parities, spin- $j$ pairs, in particular abundant among light-flavor mesons with masses around and above 2000 MeV,   would behave as chiral fields and could be of interest, among others, in strong processes where the chiral symmetry has been restored from the spontaneously broken Goldstone--, to the manifest Wigner-Weyl mode.

It is to be noticed that our method can be extended \cite{IJMPE} toward the high-spins carried by the two-spin valued representation spaces of the type 
$D^{(1/2, (n-1)/2)\oplus ((n-1)/2,1/2)}$ in (\ref{Gl1}), which are beyond the reach of the Bargmann-Wigner method, and in which case the second order wave equation is not obtained from the mass shell-condition alone but from  another covariant  projector, ${\mathcal P}_{{\mathcal W}^2}^{(j,m)}=(-{\mathcal W}^2/m^2 -j(j-1)p^2/m^2)/(2j)$, which fixes besides the mass, also the highest of the two spin  degrees of freedom. Here, ${\mathcal  W}^2$ is the squared Pauli-Lubanski operator, the second Casimir invariant of the algebra of  inhomogeneous spin-Lorentz group.
Such a projector technique, however in the bases of Lorentz-tensors,  has been originated by Aurilia and Umezawa in \cite{Umezawa} at the free spin-$3/2$ particle level, and employed  independently
in \cite{Napsuciale:2006wr} at the interacting level. In the latter work the second order differential equation resulting  from ${\mathcal P}_{{\mathcal W}^2}^{(3/2,m)}$ has been extended 
to include the  most general terms allowed by relativity and containing $\left[\partial_\mu, \partial_\nu \right]$ commutators. Such terms,  identically vanishing at the free particle level, provide upon gauging essential contributions proportional to the electromagnetic field strength tensor, $F_{\mu\nu}$ and guarantee that the resulting wave equation is  free from the Velo-Zwanziger problem \cite{VeloZw} for a $g$ factor taking  the value of $g=2$.
As long as the  Lorentz-tensors can be equivalently re-written to tensor-spinors along the prescription encoded by the above equations (\ref{inst1})-(\ref{EH_Fmunu_map}), we conclude that the technique used in \cite{Umezawa}, \cite{Napsuciale:2006wr} can be transcribed to the tensor-spinor level meaning that
the general tensor-spinors considered here can be employed in the description of spin-$j$ residing in $D^{(1/2, j-1/2)\oplus (j-1/2,1/2)}$.
Within the context of this discussion,  we believe that the relevance of our approach goes beyond its main advantage over the Bargmann-Wigner framework to correctly identify the strictly Lorentz-irreducible $D^{(j,0)\oplus (0,j)}$ degrees of freedom, while simultaneously avoiding the high order of the differential wave equations.

\section{Appendix}
\subsection{Lagrangian formulation and minimal gauging.}
In order to obtain  the master equation in (\ref{second_order_equation_spinorialtensor}) from a Lagrangian we first 
introduce the tensor-spinor, $\left[ \Gamma_{\mu\nu}^{(n/2,0)} \right]_{a_1 \dots a_{n}}^{~~~~~~~~~~~~~~~ b_1  \dots b_{n}}$, as 

\begin{equation}
\left[ \Gamma_{\mu\nu}^{(n/2,0)} \right]_{a_1 \dots a_{n}}^{~~~~~~~~~~~~~~~ b_1  \dots b_{n}}\partial^\mu \partial^\nu=\left[ \mathcal{P}^{(n/2,0)}_{F} \right]_{a_1 \dots a_{n}}^{~~~~~~~~~~~~~~~ b_1  \dots b_{n}} \partial^2, \label{Gamma_condition1}
\end{equation} 
and write down the master equation  as,

\begin{equation}
\left[ \Gamma_{\mu\nu}^{(n/2,0)} \right]_{a_1 \dots a_{n}}^{~~~~~~~~~~~~~~~ b_1  \dots b_{n}}\partial^\mu \partial^\nu  \Psi^{(n)}_{b_1  \dots b_{n}}=
-m^2\Psi^{(n)}_{a_1 \dots a_{n}} , \quad n=2j. \label{new_tensor}
\end{equation}
With this definition of $\Gamma_{\mu \nu}^{(n/2,0)}$, the free master equation (\ref{new_tensor}) can now be derived from the following Lagrangian,
\begin{equation}
\mathcal{L}_{free}^{(n/2,0)}=\left( \partial^\mu \left[ \overline{\Psi}^{(n)} \right]^{a_1  \dots a_n}  \right) 
\left[ \Gamma_{\mu\nu}^{(n/2,0)} \right]_{a_1 \dots a_{n}}^{~~~~~~~~~~~~~~~ b_1  \dots b_{n}}
\left( \partial_\nu \left[ \Psi^{(n)} \right]_{b_1  \dots b_n}  \right)
-m^2 \left[ \overline{\Psi}^{(n)} \right]^{a_1  \dots a_n} 
\left[ \Psi^{(n)} \right]_{a_1 \dots a_n}. \label{Lagrangian_free}
\end{equation}
Now the electromagnetic interaction in (\ref{second_order_equation_spinorialtensor}) and (\ref{Lagrangian_free}) 
can be introduced in the regular way through minimal gauging amounting to,
\begin{equation}
\left[ \Gamma_{\mu\nu}^{(n/2,0)} \right]_{a_1 \dots a_{n}}^{~~~~~~~~~~~~~~~ b_1  \dots b_{n}}D^\mu D^\nu \tilde{\Psi}^{(n)}_{b_1  \dots b_{n}}=-m^2\tilde{\Psi}^{(n)}_{a_1
  \dots a_{n}}, \quad D^\mu=\partial^\mu +ieA^\mu, \label{gauged_equation}
\end{equation}

and
\begin{equation}
\mathcal{L}^{(n/2,0)}=\left( D^{\ast\mu} \left[ \overline{\widetilde \Psi}^{(n)} \right]^{a_1  \dots a_n}  \right) 
\left[ \Gamma_{\mu\nu}^{(n/2,0)} \right]_{a_1 \dots a_{n}}^{~~~~~~~~~~~~~~~ b_1 \dots b_{n}}
\left( D_\nu \left[ \widetilde \Psi^{(n)} \right]_{b_1  \dots b_n}  \right)
-m^2 \left[ \overline{\widetilde \Psi}^{(n)} \right]^{a_1  \dots a_n} 
\left[ \widetilde \Psi^{(n)} \right]_{a_1  \dots a_n},
\end{equation}
respectively, where $\tilde{\Psi}^{(n)}_{b_1 \dots b_{n}}$ stands for the gauged solutions. 
To guarantee  that the gauged solutions continue being eigenstates of the Lorentz projector, i.e. that they continue transforming according to the $(j,0)\oplus(0,j)$ representation space, the tensor-spinor $\left[ \Gamma_{\mu\nu}^{(n/2,0)} \right]_{a_1 \dots a_{n}}^{~~~~~~~~~~~~~~~ b_1  \dots b_{n}}$ has to satisfy the following condition,
\begin{equation}
\left[ \mathcal{P}_{F}^{(n/2,0)} \right]_{a_1 \dots a_{n}}^{~~~~~~~~~~~~~~~ c_1  \dots c_{n}}\left[ \Gamma_{\mu\nu}^{(n/2,0)} \right]_{c_1 \dots c_{n}}^{~~~~~~~~~~~~~~~ b_1  \dots b_{n}}=\left[ \Gamma_{\mu\nu}^{(n/2,0)} \right]_{a_1 \dots a_{n}}^{~~~~~~~~~~~~~~~ b_1  \dots b_{n}}. \label{Gamma_condition2}
\end{equation}
This  condition, however, does not fix  $\left[ \Gamma_{\mu\nu}^{(n/2,0)} \right]_{a_1 \dots a_{n}}^{~~~~~~~~~~~~~~~ b_1  \dots b_{n}}$ in an unique way and one is still left with a considerable freedom in the choice of this tensor. 
In the following subsection we provide  the causality and hyperbolicity proof of the gauged equation (\ref{gauged_equation})
on the grounds of the equation (\ref{Gamma_condition1}) alone and without making any particular choice for $\left[ \Gamma_{\mu\nu}^{(n/2,0)} \right]_{a_1 \dots a_{n}}^{~~~~~~~~~~~~~~~ b_1  \dots b_{n}}$ .

\subsection{Causal propagation of the classical wave fronts of the gauged equation}
In requiring the gauged solutions of the equation (\ref{gauged_equation}) to remain in $D^{(n/2,0)\oplus(0,n/2)}$,  
we seek to expand them (in any Lorentz frame) in the basis of any complete set spanning this representation space. 
Specifically for the second-rank tensor,  $\tilde{\Psi}^{(2)}{}_{a b}(x)$, one can choose as such a set 
the six independent tensors listed  in (\ref{dof_we}), taken in the rest frame, i.e. $\left[w^{(2 )}_{\pm}(\mathbf{0},\lambda)\right]_{ab}$ .
In so doing, the corresponding gauged ${\widetilde \Psi}^{(2)}_{ab}(x)$ can be represented as
\begin{equation}
{\widetilde \Psi}^{(2)}_{ab}(x)=\sum_{\lambda, \tau } a_\tau (x, \lambda ) \left[ { w}_{\tau }^{(2)}(\mathbf{0},\lambda)\right]_{a b}, 
\quad \lambda =-1,0,+1, \quad \tau=+, -.
\label{cause1}
\end{equation}
In the case of a rank-$n$ tensor-spinor, the latter equation  generalizes to,  
\begin{equation}
\tilde{\Psi}^{(n)}_{a_1  \dots a_{n}}(x)=\sum_{\lambda, \tau }  a_{\tau }(x,\lambda)\left[w^{(n)}_{\tau}(\mathbf{0},\lambda) \right]_{a_1  \dots \alpha_{n}}, \quad \lambda=-\frac{n}{2}, ... +\frac{n}{2}, \quad \tau =+,-, \label{gauged_sol_decomposition}
\end{equation}
where $\left[w^{(n)}_{\tau}(\mathbf{0},\lambda) \right]_{a_1  \dots a_{n}}$ stands for the $2(2\frac{n}{2}+1)$ degrees of freedom spanning the
$D^{(n/2,0)\oplus (0, n/2)}$ representation space at rest.
Substituting (\ref{gauged_sol_decomposition}) in (\ref{gauged_equation}), and multiplying from the left by any one of the  conjugate states, 
$\overline{w}^{(n)}_{\pm}(\mathbf{0},\lambda)$, we obtain the following system of second order partial differential equations for the coefficients $a_{\pm}(x,\lambda)$ in (\ref{gauged_sol_decomposition}),
\begin{equation}
\left[ \overline{w}^{(n)}_{\pm}(\mathbf{0},\lambda^\prime) \right]^{a_1  \dots a_{n}}
\left[ \Gamma_{\mu\nu}^{(n/2,0)} \right]_{a_1 \dots a_{n}}^{~~~~~~~~~~~~~~~ b_1  \dots b_{n}}D^\mu D^\nu \tilde{\Psi}^{(n)}_{b_1  \dots b_{n}}=-m^2  a_{\pm}(x,\lambda^\prime). \label{system_coefficients}
\end{equation} 
In order to prove the causality and hyperbolicity of (\ref{system_coefficients}), we employ the Courant-Hilbert criterion \cite{VeloZw} which requires one to calculate the characteristic determinant. The latter is found by replacing the highest order derivative by the components of the vector $n^\mu$, the normal to the characteristic surfaces, and characterizing  the propagation of the wave fronts of the solutions to the gauged equation. 
If the characteristic determinant vanishes for real-valued time-like components, $n^0$, then the equation is hyperbolic. If in addition, 
also  $n^\mu n_\mu =0$ holds valid, then the equation is causal. 
In now applying  the Courant-Hilbert criterion to (\ref{system_coefficients}), we replace there the partial derivatives $\partial^\mu$ by the vector $n^\mu$ and arrive at,
\begin{eqnarray} \nonumber
\left[ \overline{w}^{(n)}_{\pm}(\mathbf{0},\lambda^\prime) \right]^{a_1  \dots a_{n}}
\left[ \Gamma_{\mu\nu}^{(n/2,0)} \right]_{a_1 \dots a_{n}}^{~~~~~~~~~~~~~~~ b_1  \dots b_{n}}n^\mu n^\nu \left[ w^{(n)}_{\pm}(\mathbf{0},\lambda) \right]_{b_1 b_2 \dots b_{n}}&=& \\ \nonumber
\left[ \overline{w}^{(n)}_{\pm}(\mathbf{0},\lambda^\prime) \right]^{a_1  \dots a_{n}}
\left[ \mathcal{P}_F^{(n/2,0)} \right]_{a_1 \dots a_{n}}^{~~~~~~~~~~~~~~~ b_1  \dots b_{n}}n^2 \left[ w^{(n)}_{\pm}(\mathbf{0},\lambda) \right]_{b_1  \dots b_{n}}&=& \\ \nonumber
n^2\left[ \overline{w}^{(n)}_{\pm}(\mathbf{0},\lambda^\prime) \right]^{a_1  \dots a_{n}}
\left[ w^{(n)}_{\pm}(\mathbf{0},\lambda) \right]_{a_1  \dots \alpha_{n}}&=& 
(\pm 1) n^2 \delta_{\lambda^\prime \lambda }\delta_{\pm \pm}. \label{elements_characteristic_determinant}
\end{eqnarray}
In this way, a diagonal characteristic determinant is obtained whose vanishing requires  $n^2=0$, meaning that our gauged equation is both causal and hyperbolic.


\begin{thebibliography}{99}

\bibitem{Gravity} J.\ Adamek, D.\ Daverio, R.\ Durrer, and M.\ Kunz,
{\it General relativity and cosmic formation}, Nature physics, {\bf 12},  346 (2016).

\bibitem{rot_BH}A.\ Campoleoni, H.\ A.\ Gonzalez, B.\ Oblak, and M.\ Riegler,
{\it  Rotating Higher Spin partition functions and extended BMS symmetries},
JHEP {\bf 04}, 34 (2016).
E-print Archive: arXiv:1512.03353 [hep-th].

\bibitem{A} M.\ Henneaux, G.\ L.\ G\'omez, and R.\ Rahman,
{\it Gravitational interactions of Higher-spin fermions},
JHEP 1401,  087 (2014).

\bibitem{HSReview} Rakibur Rahman, {\it High Spin Theory-Part I}, PoS {\bf Modave VIII} (2012) 004;
E-Print Archive: arXiv:1307.3199[hep-th]. 

\bibitem{BR} A.\ Barut and R.\ Raczka, {\it Group representation theory and its applications}, Vol. 2,
(Moscow, Mir, 1980) (in Russian).

\bibitem{PF} M.\ Fierz and W.\ Pauli, {\it On relativistic wave equations for particles of arbitrary spin in an electromagnetic field,\/}
Proc.\ Roy.\ Soc.\ Lond.\ A {\bf 173}, 211 (1939).

\bibitem{BW1948} V.\ Bargmann and E.\ D.\ Wigner, {\it Group theoretical discussion of relativistic wave equations},
Proc.\ Nat.\ Acad.\ Sci.\ {\bf 34}, 211 (1948).

\bibitem{Spinorbook} Moshe Carmelli and Shimon Malin, {\it Theory of spinors: An introduction,} (World Scientific, Singapore, 2000).
\bibitem{deBroigle} L.\ de Broglie, {\it Theorie general des particles a spin} (Gauthier-Villars, Paris, 1942).
\bibitem{WenliangLi} Wenliang LI, {\it A unifying framework for ghost-free Lorentz-invariant Lagrangian field theory}.
E-print Archive:  arXiv:1508.03247[gr-qc]\, .



\bibitem{TaiJunChen} Tai-jun Chen {\it et al}, {\it Higher derivatives theories with constraints:exorcising Ostrogradski's ghost},
JCAP 02, 042 (2013).
\bibitem{Joos} H.\ Joos,   {\it Zur Darstellung der inhomogenen Lorentzgruppe als Grundlage quantenmechanischer Kinematik}, 
 Fortschritte der Physik, {\bf 10},  65 (1962).

\bibitem{Weinberg} S.\ Weinberg, {\it Feynman rules for any spin},  Phys.\ Rev.\  B {\bf 133}, 1318 (1964).

\bibitem{Sankar} A.\ Sankaranarayanan, {\it $\gamma_5$ invariant constraints for any spin,} Il Nuovo Cimento   LVI A {\bf 2}, 459 (1968).

\bibitem{KimNoz} Y.\ S.\ Kim and M.\ Noz, {\it Theory and applications of the Poincar\'e group} (D. Reidel Publishing Company, Dordrecht, 1986).

\bibitem{PRD85_2012} E.\ G.\ Delgado Acosta, M.\ Kirchbach, M.\ Napsuciale, and S.\ Rodriguez, {\it Electromagnetic multipole moments of elementary spin-1/2,1, and 3/2 particles},
Phys.\ Rev.\ D {\bf 85}, 116006 (2012).

\bibitem{Greiner} W.\ Greiner, {\it Quantum Mechanics:An Introduction}, 4th edn. (Springer, Berlin, 2001) pp. 355-359.  


\bibitem{IJMPE}E.\ G.\ Delgado Acosta, V.\ M.\  Banda Guzm\'an, and M.\ Kirchbach, {\it Gyromagnetic $g_s$ factors of the spin-1/2 particles in the $(1/2^+-1/2^--3/2^-)$ triad of the four-vector spinor, $\psi_\mu$, irreducibility and linearity,}
 Int.\ J. Mod.\ Phys.\ E {\bf 24}(7),  1550060 (2015).
\bibitem{Umezawa} A.\ Aurilia and H.\ Umezawa,  {\it Projection operators in quantum relativistic fields,} Il Nuovo Cimento A {\bf 51},  14 (1967).

\bibitem{Napsuciale:2006wr} M.\ Napsuciale, M.\ Kirchbach, and  S.\ Rodriguez,
   {\it Spin-$\frac{3}{2}$ beyond the Rarita-Schwinger framework,}  Eur.\ Phys.\ J.\ A {\bf 29}, 289 (2006).


\bibitem{VeloZw} G.~Velo and D.~Zwanziger, {\it Non-causality and other defects of interaction lagrangians for particles with spin one and higher,\/ }  Phys.\ Rev.\  {\bf 188},  2218 (1969). 
\end{thebibliography}
\end{document}